\documentclass[twocolumn,showpacs,preprintnumbers,amsmath,amssymb]{revtex4}
\usepackage{graphicx}
\usepackage{dcolumn}
\usepackage{bm}

\begin{document}

\preprint{APS/123-QED}

\title{Novel ordering of an $S$ = 1/2 quasi one-dimensional\\ Ising-like anitiferromagnet
 in magnetic field}

\author{S. Kimura$^1$, T. Takeuchi$^2$, K. Okunishi$^3$, M. Hagiwara$^1$, Z. He$^4$,\\K. Kindo$^4$, T. Taniyama$^5$, M. Itoh$^5$}
\affiliation{
$^1$KYOKUGEN, Osaka University, Machikaneyama 1-3, Toyanaka 560-8531, Japan \\ $^2$Low temperature center,  Osaka University, Machikaneyama 1-1, Toyanaka 560-0043, Japan \\ $^3$Department of Physics, Niigata University, Niigata 950-2181, Japan \\ $^4$Institute for Solid State Physics, University of Tokyo, Kashiwa 277-8531, Japan\\ $^5$Materials and Structures Laboratory, Tokyo Institute of Technology, 4259 Nagatsuta, Midori Yokohama 226-8503, Japan
}

Second institution and/or address\\
This line break forced

\date{\today}

\begin{abstract}
High-field specific heat measurements on BaCo$_{\rm 2}$V$_{\rm 2}$O$_{\rm 8}$, which is a good realization of an $S$ = 1/2 quasi one-dimensional Ising-like antifferomagnet, have been performed in magnetic fields up to 12 T along the chain and at temperature down to 200 mK. We have found a new magnetic ordered state in the field-induced phase above $H_{\rm c}$ ${\simeq}$ 3.9 T. We suggest that a novel type of the incommensurate order, which has no correspondence to the classical spin system, is realized in the field-induced phase.
\end{abstract}

\maketitle
The $S$ = 1/2 one-dimensional (1D) anitiferromegnetic XXZ model is a simple spin system but exhibits very rich and exotic properties originating from quantum fluctuation inherent in the model. It is one of a few examples of many-body quantum systems, being exactly solvable by the Bethe Ansatz, and has been the object of numerous theoretical studies. An important feature of this model is a quantum critical nature of the ground state, characterized by power law decay of the correlation function in its gapless spin liquid phase~\cite{Luther,Haldane}. The quantum critical nature is a manifestation of the Tomonaga-Luttinger (TL) liquid state, which  realizes in this gapless phase. A magnetic field for the longitudinal $z$-direction can play a key role in controlling this aspect~\cite{Haldane,Bogoliubov}. The model is given by the Hamiltonian
\begin{eqnarray}
{\mathcal H} = J \sum_{\rm i} \left\{ S_{\rm i,z}S_{\rm i+1,z}+\epsilon (S_{\rm i,x}S_{\rm i+1,x}+S_{\rm i,y}S_{\rm i+1,y}) \right\} \nonumber\\
- g{\mu}_{\rm B} \sum_{\rm i}{S_{\rm i,z}}H
\end{eqnarray}
where, $J > 0$ is the antiferromagnetic exchange interaction, and $\epsilon$ $\ge$ 0 is an anisotropic parameter. The $\epsilon$-$h$ phase diagram for the ground state of the model, given by the Bethe Ansatz calculations~\cite{Bogoliubov,Yang}, is shown in Fig. 1. 
\begin{figure}
\includegraphics[width=7cm,clip]{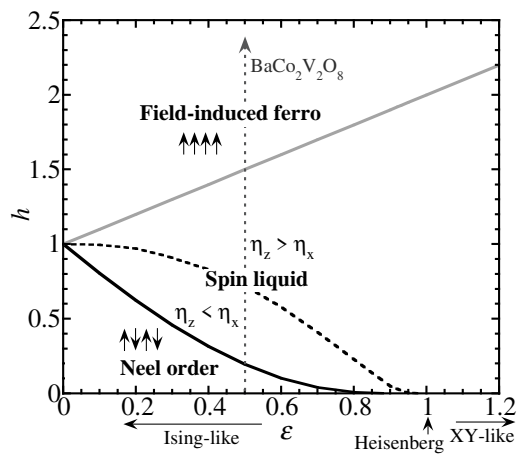}
\caption{$\epsilon$-$h$ phase diagram for 1D $S$ = 1/2 anitiferromegnetic XXZ  model. Black and gray lines show the transition field and the saturation field, respectively. Dashed line shows the field at which the crossover of the TL components occurs.}
\end{figure}
\begin{figure}
\includegraphics[width=7cm,clip]{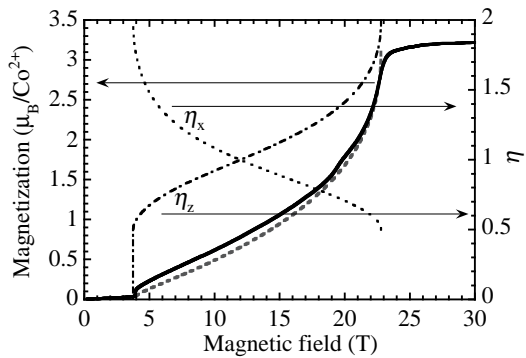}
\caption{ Magnetization curve of BaCo$_{\rm 2}$V$_{\rm 2}$O$_{\rm 8}$ and the calculated TL exponents. Solid and gray dashed lines are the experimental and theoretical magnetization curves, respectively. The Van-Vleck paramagnetic contribution is subtracted from the experimental magnetization curve. Dotted and dot-dashed lines shows the TL exponents $\eta$$_{\rm x}$ and $\eta$$_{\rm z}$, respectively}
\end{figure}
Here, $h = g{\mu}_{\rm B}H/J$ is the normalized field. Peculiar quantum critical behaviors in the fields were theoretically demonstrated in the spin liquid phase. An incommensurate correlation for the longitudinal spin component develops, in addition to a transverse staggered one~\cite{Haldane}. Asymptotic forms of the transverse and longitudinal correlation functions are expressed, respectively,  as follows:

\begin{eqnarray}
<S^{\rm x}_{\rm 0}S^{\rm x}_{\rm r}> {\simeq} (-1)^{r}r^{-{\eta}_{\rm x}}
\end{eqnarray}
and
\begin{eqnarray}
<S^{\rm z}_{\rm 0}S^{\rm z}_{\rm r}>-m^2 {\simeq} {\rm cos}(2k_{\rm F}r)r^{-{\eta}_{\rm z}}
\end{eqnarray}
where $m$ is a magnetization per spin, $\eta$$_{\rm x}$ and  $\eta$$_{\rm z}$ are the TL exponents for the $x$ and $z$ spin component, respectively, which satisfy a relation $\eta$$_{\rm x}$$\eta$$_{\rm z}$ = 1, and $k_{\rm F}$ = $\pi$(1/2-$m$). 

In this letter, we discuss the quasi-1D $S$ = 1/2 Ising-like antiferromagnet with $\epsilon$ $<$ 1 in the longitudinal fields at very low temperature. Inevitable three-dimensional (3D) interactions in real quasi-1D materials make the quantum critical state unstable, and can lead to a long range order at a finite temperature. As a result, the magnetic ordering with exotic properties, reflecting the peculiar quantum critical nature of the 1D XXZ model, is expected to appear in the quasi-1D Ising-like antiferromagnet. Figure 1 shows that a quantum phase transition from the N{\' e}el ordered phase to the quantum critical spin liquid one occurs at a critical field $H_{\rm c}$ in the Ising-like system~\cite{Haldane,Bogoliubov}. An important point for the Ising-like XXZ case
is that the incommensurate correlation is enhanced above the critical field $H_{\rm c}$,
in contrast to the usual isotropic Heisenberg model, for which the staggered correlation of the $xy$ spin components
is always dominant in a magnetic field. Then the interchain coupling brings 
a quite interesting situation in the gapless phase above $H_{\rm c}$, as will be argued later. We will report the first observation of the magnetic ordering in the field-induced phase of $S$ = 1/2 quasi-1D Ising-like antiferromagnet above $H_{\rm c}$ by specific heat measurements on BaCo$_{\rm 2}$V$_{\rm 2}$O$_{\rm 8}$.

In BaCo$_{\rm 2}$V$_{\rm 2}$O$_{\rm 8}$, a magnetic Co$^{\rm 2+}$ ion is surrounded by six oxygen atoms, and the edgeshared CoO$_{\rm 6}$ octahedra form a screw chain along the $c$-axis~\cite{Wichmann}.
 At zero field, BaCo$_{\rm 2}$V$_{\rm 2}$O$_{\rm 8}$ undergoes a long range ordering at 5.4 K~\cite{He1,He2,He3}. When the external field is applied for the $c$-axis, which is the easy axis of BaCo$_{\rm 2}$V$_{\rm 2}$O$_{\rm 8}$, a field-induced transition occurs at $H_{\rm c}$ ${\simeq}$ 3.9 T, as shown in Fig.~2~\cite{He1,He2,Kimura1,Kimura2}. In contrast to the spinflop transition of classical antiferromagnets with easy-axis anisotropy, which exhibits a linear increase of the magnetization above the transition field, the magnetization in BaCo$_{\rm 2}$V$_{\rm 2}$O$_{\rm 8}$ shows a non-linear and steep increase above $H_{\rm c}$. This behavior suggests strong quantum fluctuation in the field-induced phase. In fact, previous specific heat measurements showed that no magnetic ordering is observed down to 1.8 K in the field-induce phase for $H$ $>$ $H_{\rm c}$~\cite{He1,He2}. Therefore, the field-induced transition in the temperature region above 1.8 K causes a destruction of the long range magnetic order.~\cite{He1,He2}. In our recent study, we revealed that the above mentioned quantum phase transition of the $S$ =1/2 1D Ising-like antiferromagnet, which brings about the quantum critical nature in the field-induced phase, is responsible for the destruction of the magnetic order in BaCo$_{\rm 2}$V$_{\rm 2}$O$_{\rm 8}$~\cite{Kimura1,Kimura2}.  The magnetization curve is explained by the $S$ =1/2 1D XXZ model with $\epsilon$ ${\simeq}$ 0.5, $g$ = 6.2 and $J$/$k$$_{\rm B}$ = 65 K, as shown in Fig. 2. An advantage of BaCo$_{\rm 2}$V$_{\rm 2}$O$_{\rm 8}$ is its rather low critical field $H_{\rm c}$ ${\simeq}$ 3.9 T, that is easily accessible by commercial superconducting magnets. Thus, BaCo$_{\rm 2}$V$_{\rm 2}$O$_{\rm 8}$ is very suitable material for studies of the field-induced phase in the $S$ =1/2 quasi-1D Ising-like antiferromagnet. In this study, with further decreasing temperature by using a dilution refrigerator, we find a new long range ordered phase above $H_{\rm c}$. The transition between the N{\' e}el ordered phase to the field-induced ordered one is revealed to be of first order. From the theoretical considerations based on the Bethe Ansatz calculations, we propose that a novel type of the incommensurate order, which has no correspondence to the classical spin system, is realized in the field-induced phase.

The specific heat of  BaCo$_{\rm 2}$V$_{\rm 2}$O$_{\rm 8}$ for $H{\parallel}c$ was measured by means of a quasi-adiabatic heat pulse method in high magnetic field up to 12 T and at temperature down to 200 mK. We use a dilution refrigerator and a superconducting magnet. Magneto-caloric effect was also measured at $T$ = 1.0 and 0.5 K. Single crystals of BaCo$_{\rm 2}$V$_{\rm 2}$O$_{\rm 8}$, grown by a spontaneous nucleation method~\cite{He3}, were used for the measurements. 
\begin{figure}
\includegraphics[width=7cm,clip]{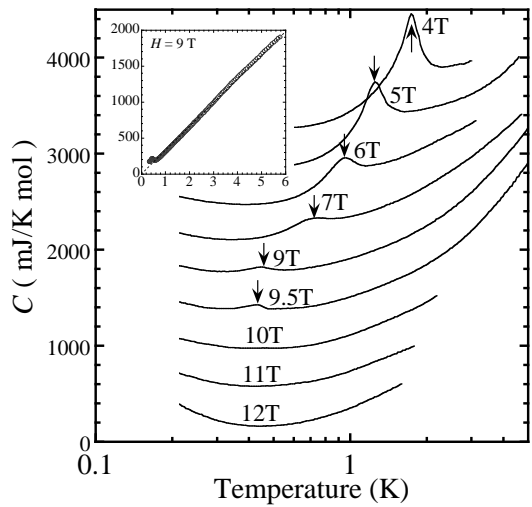}
\caption{Temperature dependence of the specific heat of BaCo$_{\rm 2}$V$_{\rm 2}$O$_{\rm 8}$ for $H$${\parallel}$$c$. Each specific heat shifts up by 400 mJ/K mol with decreasing field from 12 T. Inset shows an extended figure of the specific heat observed at 9 T. Open circles are the experimental data and dashed line is a guide for the eye.}
\end{figure}

Figure 3 shows the temperature dependence of the heat capacity observed in BaCo$_{\rm 2}$V$_{\rm 2}$O$_{\rm 8}$ for $H$${\parallel}$$c$. In the field region above  $H_{\rm c}$, a $\lambda$-peak, which corresponds to the long range magnetic order, is observed at temperature below 1.8 K. As the field is increased, the peak position shifts toward lower temperature. The $\lambda$-peak diminishes with increasing the field and disappear in the field region above 10 T. Slight upturn of the specific heat is seen in the low temperature region, presumably owing to the contribution from the nuclear magnetism. The observed ordering temperatures are plotted together with the data obtained from the previous measurements by He {\em et al.}~\cite{He1,He2} and the transition fields determined by the magnetization and magetocarolic effect measurements in Fig. 4. The inset of Fig. 4 shows the field dependence of magnetocarolic effect, observed at 1.0 and 0.5 K. Sharp peaks at $H_{\rm c}$ ${\simeq}$ 3.9 T, accompanied by a small hysteresis, exhibit a phase boundary of weak first order, as plotted by triangles in Fig. 4. Thus, our measurements reveal the appearance of a new ordered state in the field-induced phase above $H_{\rm c}$ ${\simeq}$ 3.9 T at temperature below 1.8 K. The peaks, found in the magnetocaloric effects at $H_{\rm c}$, show that the transition between the N{\' e}el ordered phase to the field-induced ordered one is accompanied by latent heat. The concave peak, which corresponds to an endothermic behavior of the sample, in the field-ascending process suggests larger entropy in the field-induced phase compared with the N{\' e}el ordered one.
\begin{figure}
\includegraphics[width=7cm,clip]{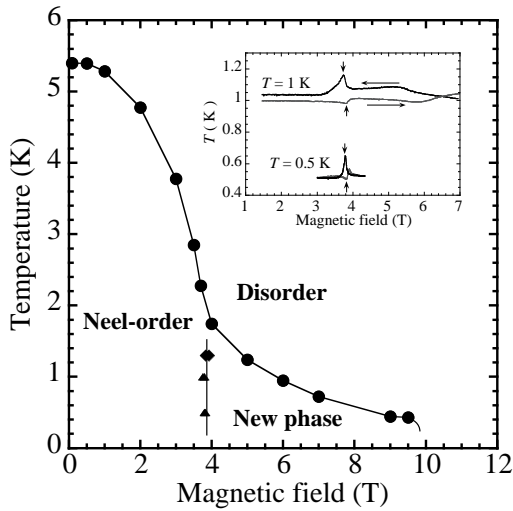}
\caption{Phase diagram of BaCo$_{\rm 2}$V$_{\rm 2}$O$_{\rm 8}$ for $H$${\parallel}$$c$. Closed circles show the transition temperatures determined from the heat capacity measurements. Diamonds and triangles show the transition fields determined from the magnetization curves and magnetocarolic effects, respectively. Solid curves are guides for the eye. Inset shows the magnetocarolic effects, observed at 1.0 K and 0.5 K. Gray and black curves show the experimental data, observed in the field-ascending and descending processes, respectively.}
\end{figure}

As we mentioned before, a gapless spin liquid state realizes in the field region above the transition field $H_{\rm c}$ for ideal 1D $S$ = 1/2 Ising-like antiferromagnets. 
The magnetic excitation in the spin liquid state is characterized by $k$-linear dispersion with soft mode at incommensurate wave vectors~\cite{Haldane}. Our previous ESR measurements on BaCo$_{\rm 2}$V$_{\rm 2}$O$_{\rm 8}$ at 1.3 K suggested the development of such incommensurate soft modes for $H$ $>$ $H_{\rm c}$~\cite{Kimura2}. A $T$-linear dependence of the specific heat above the ordering temperature, indicative of the gapless $k$-linear dispersion~\cite{Hagiwara}, further supports a realization of the gapless spin liquid state; see the inset of Fig. 3. For the Heisenberg or XY system with $\epsilon$ $\ge$ 1, the transverse staggered correlation is always dominant compared with the longitudinal one in the spin liquid phase up to the saturation field as shown in Fig. 1~\cite{Bogoliubov}. Therefore, usual N{\' e}el type ordering of the transverse spin components should take place in the quasi-1D XY or Heisenberg antiferromagnet. The quantum critical behavior of the Ising-like system, however, is quite different from that of the system with $\epsilon$ $\ge$ 1.
Dotted and dot-dashed lines in Fig. 2 show the field dependences of the TL exponents $\eta$$_{\rm x}$ and $\eta$$_{\rm z}$ for the $S$ = 1/2 1D XXZ model respectively, calculated by the Bethe Ansatz integral equation ~\cite{Bogoliubov} with $\epsilon$ = 0.46, $g$ = 6.2 and $J$/$k$$_{\rm B}$ = 62.5 K. These parameters were estimated from the analysis of the magnetization curve~\cite{Kimura1}. It should be mentioned that a small ferromagnetic next-nearest neighbor (NNN) intrachain interaction in BaCo$_{\rm 2}$V$_{\rm 2}$O$_{\rm 8}$ was suggested from our ESR measurements~\cite{Kimura2}. However, the NNN interaction is neglected in this discussion, because it is estimated to be less than ten percent of the nearest neighbor one. In contrast to the Heisenberg or XY system, $\eta$$_{\rm z}$$ < $$\eta$$_{\rm x}$, which indicates a dominance of the longitudinal incommensurate correlation compared with the transverse one, is satisfied for the Ising-like system in a certain field region above $H_{\rm c}$. When the 3D interaction is relevant, the dominant longitudinal correlation in the chain is expected to develop to the long range order for the quasi-1D material. Thus, we propose that the long range ordered phase with an incommensurate spin structure appears in BaCo$_{\rm 2}$V$_{\rm 2}$O$_{\rm 8}$ above $H_{\rm c}$. In this structure, the spins align to be colinear along the $c$-axis with modulation of those amplitude, characterized by the incommensurate wave number $k$ = 2$k_{\rm F}$. Actually, the recent theoretical study demonstrated that such an incommensurate order, which results from  the dominance of the $\eta$$_{\rm z}$, can be stabilized by the interchain molecular fields for the $S$ = 1/2 bond-alternating chain system with frustration~\cite{Maeshima}. The concave curve of the phase boundary between the ordered phase in the field-induced region and the disordered one in BaCo$_{\rm 2}$V$_{\rm 2}$O$_{\rm 8}$ is qualitatively different from convex phase boundaries found in the Heisenberg spin gap systems, such as TlCuCl$_{\rm 3}$~\cite{Oosawa}, for which the ordering of transverse spin components appears. We consider that this reflects the difference of the spin structures in the ordered phase. The appearance of the incommensurate order in the longitudinal field was discussed for the first time in the $S$ = 1 quantum 1D chain with large single-ion type easy-axis anisotropy~\cite{Sakai}. To our knowledge, however, the incommensurate order in such a system has not been confirmed experimentally yet.

In the ordered phase above $H_{\rm c}$, the incommensurate soft mode will turn into the gapless Goldstone mode as a consequence of the breaking of a quasi-continuous translation symmetry, whereas the magnetic excitation in the N{\' e}el ordered phase is the spinon with a finite gap~\cite{Ishimura}. Thermal excitation is activated in presence of the gapless mode, giving rise to the larger entropy for the field-induced phase, as suggested by the magnetocaloric effect. With further increasing the field, a crossover of the TL exponents between $\eta$$_{\rm x}$ and $\eta$$_{\rm z}$ occurs, and then the transverse correlation becomes dominant in the 1D chain, as shown in Fig. 2. The disappearance of the long range order above 10 T may be attributed to this crossover. Around the crossover field, the long range order should be suppressed, as the order of both the longitudinal and transverse spin components usually cannot coexist. Competition between these two kinds of order, which becomes significant around the crossover field because of the equivalence between the $\eta$$_{\rm x}$ and $\eta$$_{\rm z}$, disturbs the system in taking a unique stable spin structure. In the higher field region above 12 T, we also expect an appearance of another  phase, characterized by staggered order with respect to the $xy$ spin components, owing to a development of the dominant transverse correlation.

Recent quantum Monte-Carlo simulation showed that the incommensurate long range order can be stabilized only by the interchain exchange interaction for the $S$ = 1/2 quasi-1D XXZ model~\cite{Suzuki}. It should be mentioned that this incommensurate order is different from the spin density wave state of conducting linear chain systems, which is arisen by the nesting of Fermi surface, and is caused entirely by the quantum fluctuation inherent in the gapless spin liquid state of the $S$ = 1/2 1D XXZ model. It is theoretically known that the $S$ = 1/2 1D XXZ model can be represented by pseudofermion~\cite{Leib,Bulaevskii}. The field-induced quantum phase transition in the $S$ = 1/2 1D Ising-like antiferromagnet is described as that from the charge ordered phase with alternative alignments of the hole and pseudofermion to the TL liquid one with an incommensurate charge density wave (CDW) correlation of the pseudofermion. The incommensurate order of the longitudinal spin component in the $S$ = 1/2 1D XXZ model is a spin version of the CDW order for the pseudofermion. Such a CDW order, corresponding to the incommensurate order for the spin chain, is likely assisted by the spin-lattice coupling~\cite{Emery}. If the spin-lattice coupling is relevant, similar behaviors to the spin-Peierls system in the magnetic field~\cite{Pytte,Cross} are expected for the ordering in the field-induced phase of BaCo$_{\rm 2}$V$_{\rm 2}$O$_{\rm 8}$. As the orbital degeneracy of Co$^{\rm 2+}$ is not lifted by octahedral cubic ligand field~\cite{Abragam,Lines}, relatively strong spin-lattice coupling is anticipated in BaCo$_{\rm 2}$V$_{\rm 2}$O$_{\rm 8}$. It is yet unclear that the spin-lattice coupling affects the ordering of BaCo$_{\rm 2}$V$_{\rm 2}$O$_{\rm 8}$ in the field-induced phase. For more detailed studies of the ordered state in BaCo$_{\rm 2}$V$_{\rm 2}$O$_{\rm 8}$ above $H_{\rm c}$, direct observations of the spin and lattice structures by neutron scattering measurements are desired.

The authors are grateful to Dr. T. Suzuki for valuable discussions. This work was partly supported by Grant-in-Aid for Young Scientists (B) (No. 18740183 and No. 1870230) and by Grant-in-Aid for Science Research on Priority Areas "High Field Spin Science in 100 T" (No. 451) from the Ministry of Education, Science, Sports, Culture and Technology (MEXT) of Japan.


\end{document}